\documentclass[]{aa}
\usepackage{graphicx}

\begin{document}
  
   \title{Properties of the background of EPIC-pn onboard XMM-Newton}
 
   \author{H. Katayama\inst{1}, I. Takahashi\inst{2}, Y. Ikebe\inst{3},
K. Matsushita\inst{3}, and M. J. Freyberg\inst{3}}

   \institute{Department of Earth and Space Science, Graduate School of
   Science, Osaka University, 1-1 Machikaneyama, Toyonaka, 560-0043 Osaka,
   Japan\\
   \email{hkatayam@ess.sci.osaka-u.ac.jp}
   \and
   Department of Physics, University of Tokyo, 7-3-1 Hongo, Bunkyo-ku,
   113-0033 Tokyo, Japan
   \and
   Max-Planck-Institut f\"ur Extraterrestrische Physik, Postfach 1312, 85741
   Garching, Germany
   } 

   \date{Received ; accepted }

   \abstract{ We have investigated the background properties of EPIC-pn
   onboard XMM-Newton to establish the background subtraction
   method. Count rates of the background vary violently by two orders of
   magnitude at the maximum, while during the most quiet period, these
   are stable within 8 \% at a 1 $\sigma$ level. The overall spectrum is
   dominated by particle events above 5 keV, and its spatial variation
   is also found.
   The long-term variation of the background is also investigated with
   CAL CLOSED data,
   where the filter wheel was in closed position with the internal
   calibration source illuminating the sensitive area.
   The average background count rate decreased by 20 \% from
   March 2000 to January 2001, but it regained in February 2001.
   For the modeling of the background spectrum, we investigate
   relations between the 2.0--7.0 keV count rate and some characteristic
   parameters. The 2.0--7.0 keV background count rate shows a good
   correlation with the count rate of events outside the field of view. 
   This correlation is usable for the modeling of the
   background.  
   \keywords{ X-rays: general -- 
              Instrumentation: detectors
               }
   }
   \authorrunning{H. Katayama et al.}
   \titlerunning{Background of EPIC-pn.}

   \maketitle

\section{Introduction}

In X-ray data analysis, understanding properties of background in the
data is crucial in order to extract actual source X-rays, in particular
for extended objects with low surface brightness like galaxy clusters
and supernova remnants (SNRs) having low signal to noise ratios. The
most sophisticated high quality X-ray detectors currently used in X-ray
observations in orbit are X-ray CCDs, which was for the first time on
board ASCA (1993--2000; Tanaka et al. \cite{tanaka94}) and are currently
utilized on XMM-Newton (Jansen et al. \cite{jansen}) and Chandra
(Weisskopf et al. \cite{weisskopf}).  The background of ASCA/SIS
(Solid-state Imaging Spectrometer) is rather stable and its count rate
correlates with the local cutoff rigidity value, indicating that the
background is mainly induced by high energy particles, except at the South
Atlantic Anomaly (SAA) (see e.g. Gendreau \cite{gendreau95a}; Ueda
\cite{ueda}).  On the other hand, the X-ray CCD cameras onboard
XMM-Newton and Chandra exhibit violent variability as large as two orders
of magnitudes (Markevitch \cite{markevitch}; Lumb \cite{lumb}).  The responsible
difference between ASCA and the currently working two satellites is
their orbits.  ASCA took an almost circular orbit with an average
altitude of 520--620 km, while XMM-Newton and Chandra take an highly
elliptical orbit with an apogee of about 100,000 km and a perigee of
about 10,000 km and thus mostly fly outside the Earth's
magnetosphere. The rapid variation is thought to be caused by protons
scattered through the mirror system (often referred to as ``soft protons'').
In particular, protons in the energy of 100--200 keV degrade the
performance of Chandra ACIS-I (Prigozhin et al. \cite{prigozhin}).  Lumb
(\cite{lumb}) suggests that the rapid flux variation of those soft
protons is caused by magnetic reconnections rather than solar flares.

XMM-Newton, with the largest effective area of the X-ray telescope and
the high quantum efficiency of the CCD cameras, should be the best
suitable observatory for low surface brightness objects.  In this paper,
we show the properties of the background of the EPIC-pn onboard
XMM-Newton, following reports by Briel et al (\cite{briel}), Freyberg et al.
(\cite{freyberga},\cite{freybergb}), and Lumb (\cite{lumb}).  We also study correlations
between the background and some characteristic parameters to look for a
possible method of the background modeling.

\section{Data and Screening}

We used EPIC-pn data taken during the calibration and performance
verification (Cal-PV) phase.  We selected  data sets  taken in  full frame
mode  with  thin1 filter, excluding data with bright sources and observation
time longer than 15 ksec to ensure enough count statistcs.
In addition to those data sets, we also use a data set taken with the
filter wheel in the closed position (hereafter referred as CLOSED),
which includes no contributions from photons or particles through the
X-ray telescope.  The data sets used in our analysis are summarized in
Table \ref{tbl:ms_tbl1}.

\begin{table*}
\begin{center}
  \caption{PN data summary}
  \label{tbl:ms_tbl1}
  \begin{tabular}{l l l l}\hline
\multicolumn{3}{c}{Filter Thin} \\ \hline\hline
  Obs ID                        & Exposure[s]$^a$ &   Object  & symbols$^b$ \\ \hline
  0063\_0123100201\_PNS001      & 11411   & MS0737.9+7441    & closed circle  \\
  0070\_0123700101\_PNS003      & 27695   & Lockman Hole     & closed square  \\ 
  0071\_0123700201\_PNS003      & 21009   & Lockman Hole     & closed triangle  \\ 
  0073\_0123700401\_PNS003      & 9977    & Lockman Hole     & open circle \\ 
  0078\_0124100101\_PNS003      & 21403   & RXJ0720.4-3125   & open square \\
  0081\_0123701001\_PNS003      & 17569   & Lockman Hole     & open triangle\\ 
  0082\_0124900101\_PNS003      & 17665   & MS1229.2+6430    & open diamnod \\
  0086\_0125300101\_PNS013      & 26210   & J104433.04-012502.2 & closed
   asterisk\\
  0181\_0098810101\_PNS003      & 18529   & WW Hor           &open asterisk\\ \hline
\multicolumn{3}{c}{} \\
\multicolumn{3}{c}{CLOSED} \\ \hline\hline
  Obs ID                     & Exposure[s]$^a$   &   Object   \\ \hline
0059\_0122320701\_PNS003     &  42258   & \\ \hline
 \end{tabular}
\end{center} 
$^a$:  after the screening of  the flaring component \\
$^b$:  symbols  in Figure 4 (left), 5, 11, 12, and 13.
\end{table*}

\begin{figure}
\resizebox{\hsize}{!}{\includegraphics{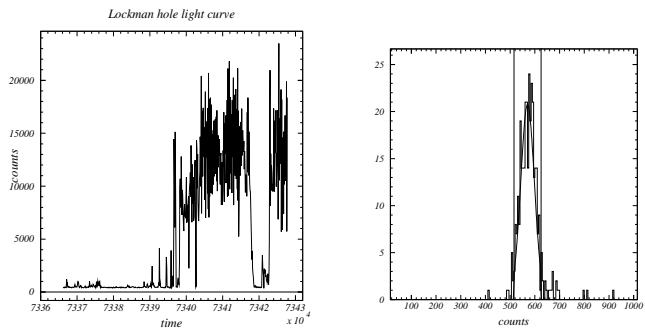}}
\caption{(Left) Light curve of a Lockman Hole observation. (Right) Count 
 rate histogram of the light curve. Two vertical lines are threshold of
 $\pm 2\sigma$.}
\label{fig:ms_fig1}
\end{figure}

\begin{figure}
\resizebox{\hsize}{!}{\includegraphics{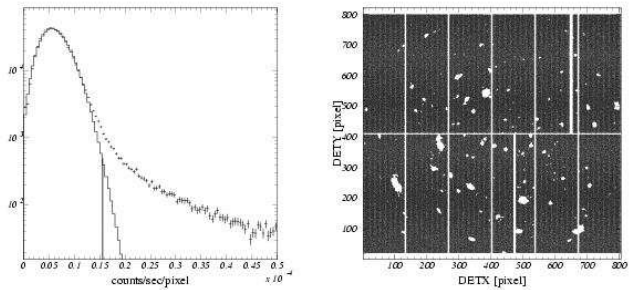}}
\caption{(Left) Pixel-counts histogram. The histogram of count rate of each
image bin from the smoothed image. Solid line represents the best fit
 of Poisson distribution. The vertical line is threshold of 2.5 times
 the average. (Right) Image masked pixels exceed threshold. }
\label{fig:ms_fig2}
\end{figure}

We have selected single and double pattern events.
Fig. \ref{fig:ms_fig1} shows a light curve from one of the Lockman hole
observations. The light curve mainly consists of two characteristic
components. One is the flaring component that is characterized by strong
and rapid variability by two orders of magnitude at the maximum. The
other part of the light curve shows rather stable count rates and ought
to be used for scientific analysis. In order to exclude the flaring
component, time periods where the count rate deviates from the mean
value during the quiescent period by $\pm 2\sigma$ are
excluded. Celestial sources and noisy CCD pixels are also excluded from
the data in the following way. The image was binned with 40 sky pixels
($2''$) and smoothed by a Gaussian of $\sigma = 4''$. From the smoothed
image made, we then made a pixel-counts histogram, which is the
histogram of the count rate of each image bin
(Fig. \ref{fig:ms_fig2}). The pixel-counts histogram was fitted with a
Poisson distribution. The pixels with count rates exceeding 2.5 times the
average value were excluded as a source or noisy pixels. After
eliminating those high count pixels, we again made a light curve and
performed the temporal screening as described above.  The average
exposure time remaining after the screening is about 19 ksec. The sample
may represent typical observations. In the following sections, we
investigate  stability of the  energy spectrum and the spatial
distribution of the background.

\section{Spectrum of EPIC-pn background}

\begin{figure}
\resizebox{\hsize}{!}{\includegraphics{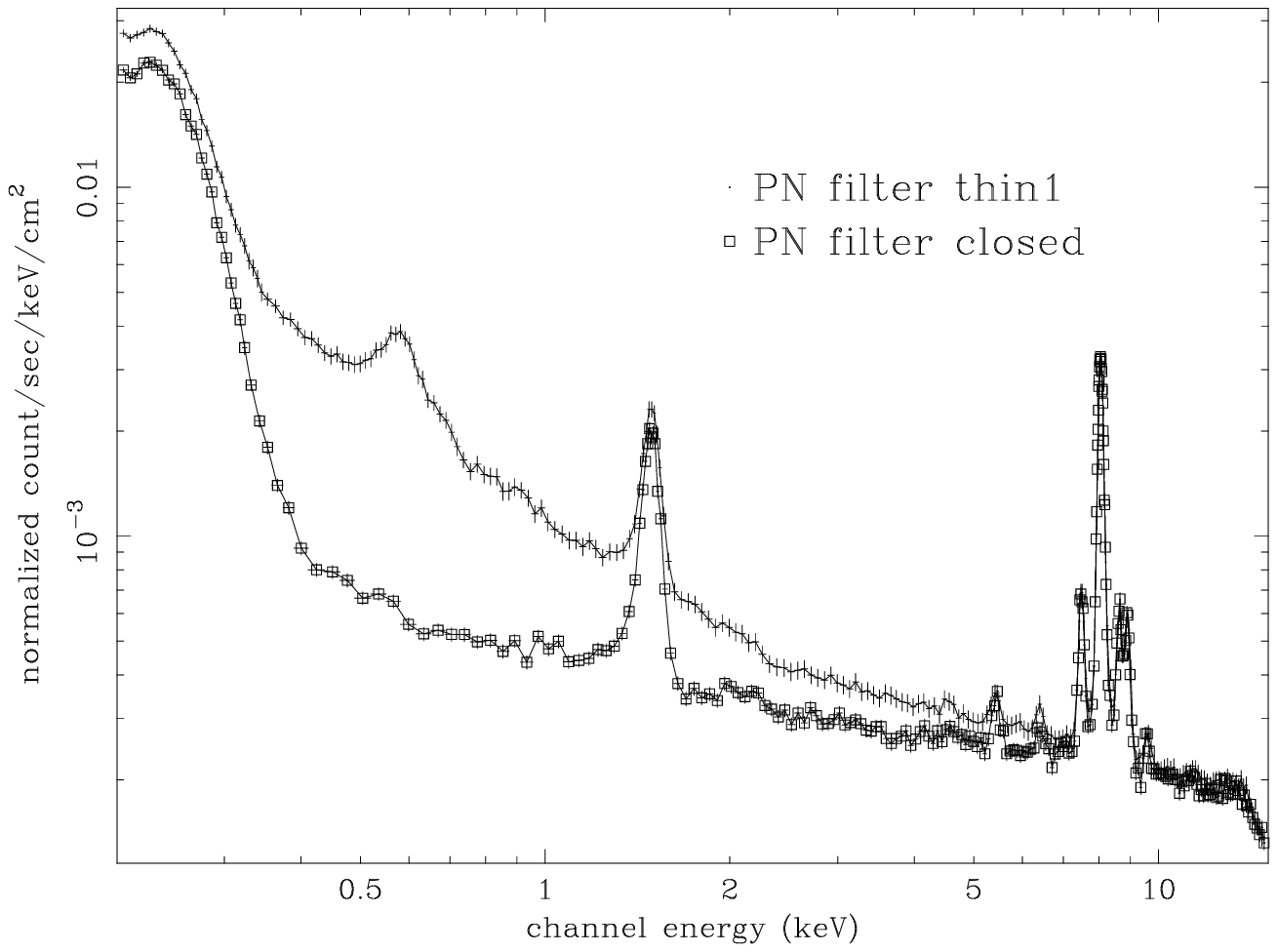}} 
\caption{Average (dot) and internal (open square) background spectra of
the EPIC-pn. The prominent line features are Al-K (1.49keV), Ni-K (7.48
keV), Cu-K (8.05,8.91 keV), and Zn-K (8.64,9.57keV) fluorescence
lines. }
\label{fig:ms_fig3}
\end{figure}

 Figure \ref{fig:ms_fig3} shows the average background spectrum of EPIC-pn
obtained from observations of all the thin filter data listed in Table
\ref{tbl:ms_tbl1}. Compared with the energy spectrum from the CLOSED
data, which is overlaid in the same figure, the background spectrum is
found to be dominated by the particle origin events above 5 keV, while,
below 5 keV, the  cosmic X-ray background (CXB) is the major
contribution. Below 0.3 keV, another non-X-ray noise emerges.  The
prominent line features are Al-K (1.49keV), Ni-K (7.48 keV), Cu-K
(8.05,8.91 keV), and Zn-K (8.64,9.57keV) lines (Freyberg et al. \cite{freyberga}). The excess emission at
0.6 keV is very likely due to characteristic emission lines from O VII
and O VIII that associated with the local bubble (e.g. Tanaka \& Bleeker
\cite{tanaka77}, Gendreau et al. \cite{gendreau95b})

\begin{figure*}
\includegraphics[width=17cm]{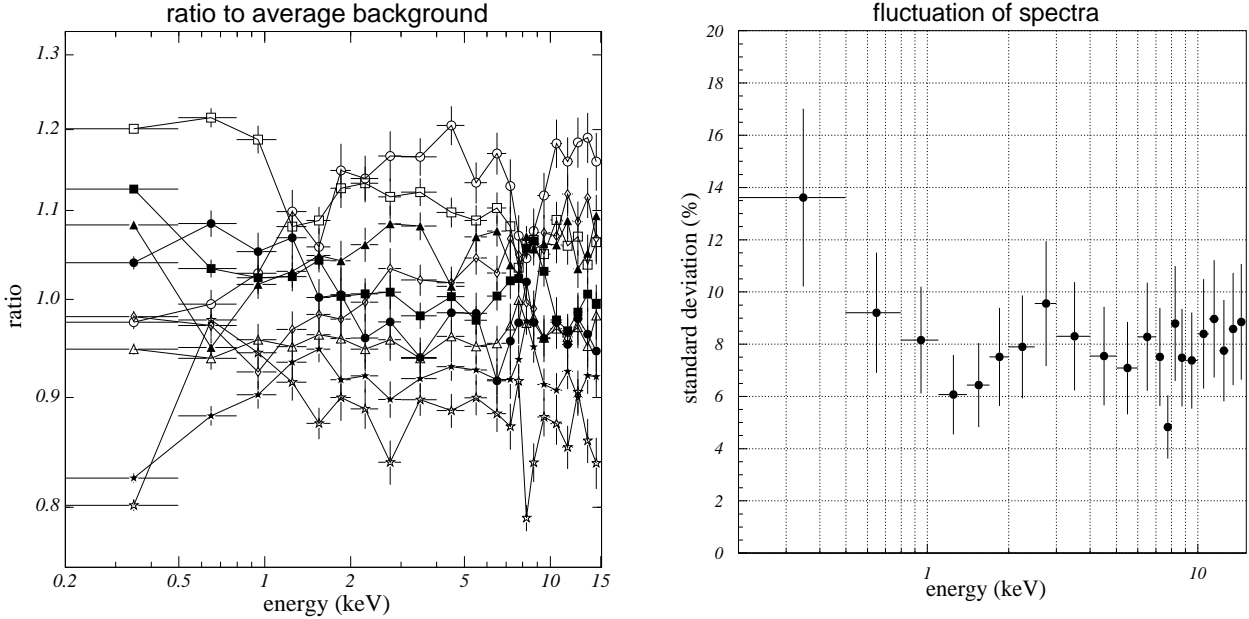}
\caption{(Left) Ratios of individual background spectrum to the
 average.  The meaning of the symbols are summarized in Table 1.
(Right) Standard deviation of background spectra in each energy
 band at a 1 $\sigma$ level.}
\label{fig:ms_fig4}
\end{figure*}

We next accumulated energy spectrum from each data set and obtained its
ratio to the average spectrum, which are shown in the left panel of Fig. 
\ref{fig:ms_fig4}.  The variations of count rate in each energy band are
quantified with the standard deviation at a 1 $\sigma$ and illustrated
in the right panel of Fig. \ref{fig:ms_fig4}.  In the 0.5--12 keV, it is
almost constant at $\sim$ 8\%, while below 0.4 keV it increases to
$\sim$ 14\%.  The variations at Al and Cu lines are significantly
smaller than those of continuum.
These lines come from the electronic board below the CCD and do not
affected by the soft protons,  since those which can be scattered through the
mirror system can not penetrate the CCD.
Therefore, high energy particles which induce these lines may have
a smaller variability than the soft protons.

\begin{figure}
\resizebox{\hsize}{!}{\includegraphics{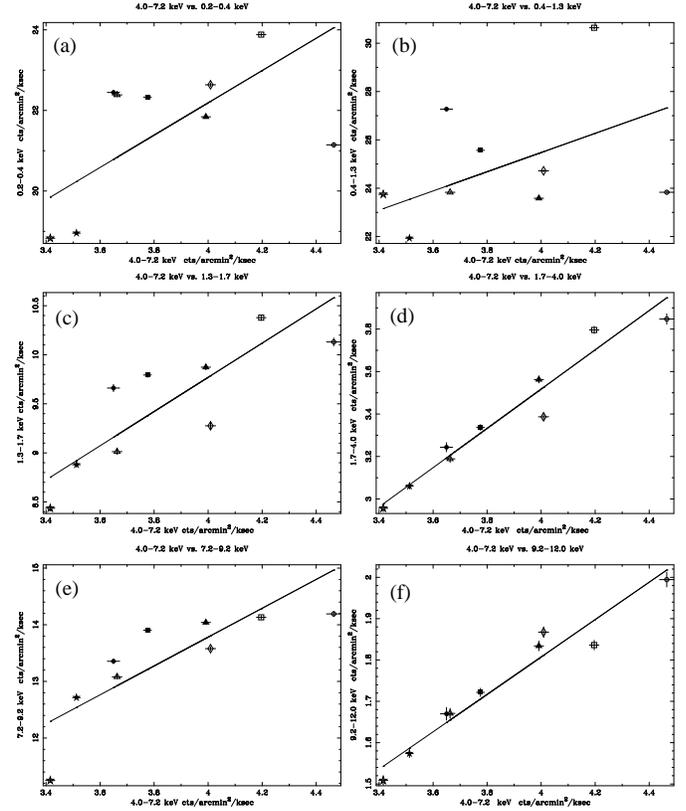}} 
\caption{Relations of background count rates between 4.0--7.2 keV and
 other six energy bands; (a) 0.2--0.4 keV, (b) 0.4--1.3 keV, (c)
 1.3--1.7 keV, (d) 1.7--4.0 keV, (e) 7.2--9.2 keV, and (f) 9.2--12.0
 keV.
The meaning of the symbols are summarized in Table 1.
The solid lines correspond to the best fit regression lines.}  
\label{fig:ms_fig5}
\end{figure}

In order to investigate the stability of the shape of the background
spectrum, we divided the spectrum into seven energy bands (0.2--0.4 keV,
0.4--1.3 keV, 1.3--1.7 keV, 1.7--4.0 keV, 4.0--7.2 keV, 7.2--9.2 keV,
and 9.2--12.0 keV) and derived the ratios between different energy
bands. Figure \ref{fig:ms_fig5} shows the relations of count rates between
4.0--7.2 keV and the other six energy bands. In the 1.3--12.0 keV band, the shape
of the background spectrum is stable within 5 \%. Below 1.3 keV, no good
correlation is seen. In the 0.4--1.3 keV band, this is probably due to
variation of the CXB component in the soft energy region.  No
correlation seen between the 0.2--0.4 keV and 4.0--7.2 keV indicates
that the non-X-ray noise that dominates the 0.2--0.4 keV count rate vary
independently from the CXB and the particle origin events.

\section{Spatial distribution of background}

\begin{figure}
\resizebox{\hsize}{!}{\includegraphics{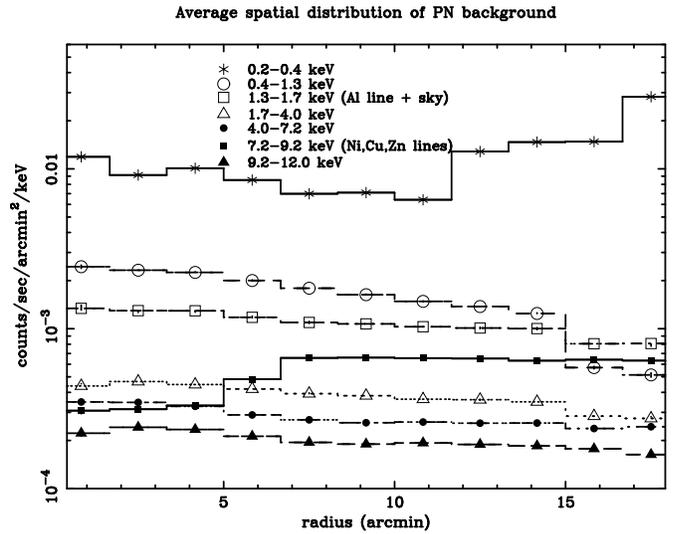}}
\caption{Radial count rate profile of EPIC-pn background.}  
\label{fig:ms_fig6}
\end{figure}

Fig. \ref{fig:ms_fig6} displays the radial count rate profile of the
backgrounds in different energy bands.
 In the 0.2--0.4 keV band profile,
 the count rate increases beyond $\sim$12 arcmin, which caused
by a concentration of the non-Xray noise around the readout node.
In the 0.4--4.0 keV band where CXB dominates, the profiles show the
telescope vignetting effect.  A sudden drop at 15 arcmin radius
corresponds to the edge of the field of view (FOV), outside which no sky
X-rays are supposed to be detected.  The 7.2--9.0 keV band is dominated
by Cu line, which is originated from the electronics circuit board
placed beneath the CCD except the central about 6 arcmin region (see
 Freyberg et al. (\cite{freybergb})) In the highest
energy band, 9.2--12.0 keV, the count rate profile is almost flat with
some excess in the central $\sim6$ arcmin region, which may correspond
to the structure of the electronics circuit board.

\begin{figure}
\resizebox{\hsize}{!}{\includegraphics{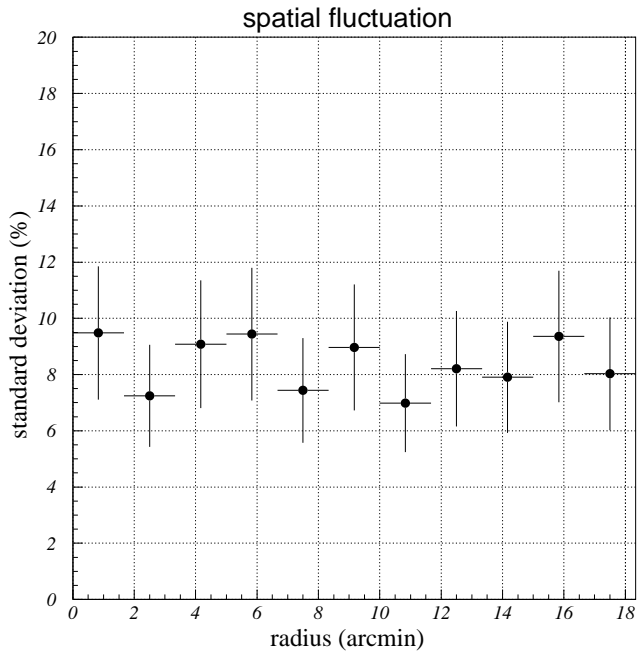}}
\caption{Standard deviation of the Radial profile of EPIC-pn background.}  
\label{fig:ms_fig7}
\end{figure}

\begin{figure*}
\includegraphics[width=17cm]{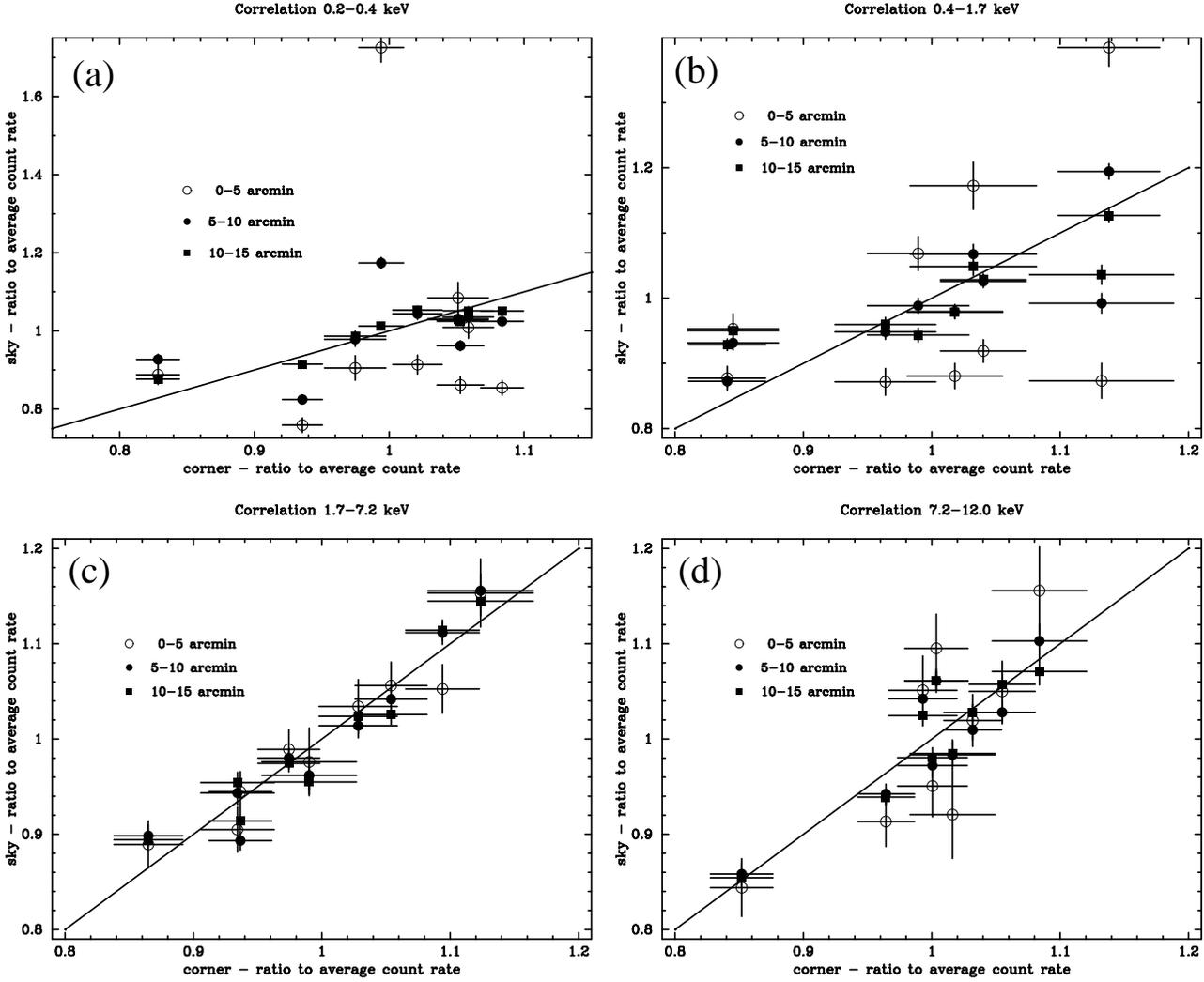} 
\caption{Correlations of count rate between the region of outside of FOV (corner)
and inside of FOV (sky) in different energy bands; (a) 0.2--0.4 keV, (b)
0.4--1.7 keV, (c) 1.7--7.2 keV, and (d) 7.2--12.0 keV. 
 Each count rate is normalized by the average count rate.
Open circles, filled circles, and filled squares indicate the count rate in the
$r=0$--5$'$, 5--10$'$, and 10--15$'$, respectively.
The solid lines correspond to the best fit regression lines.}
\label{fig:ms_fig8}
\end{figure*}

As for the energy spectrum, we examined the stability of the radial
count rate profile.  Fig. \ref{fig:ms_fig7} shows the 1 $\sigma$
standard deviation in the 2.0--7.0 keV band, which is virtually constant
over the detector.  Fig. \ref{fig:ms_fig8} shows the correlations of
count rates between outside of FOV and inside of FOV for different
energy bands (0.2--0.4 keV, 0.4--1.7 keV, 1.7--7.2 keV, and 7.2--12.0
keV).   Open circles, filled circles, and filled squares indicate the
count rate in the $r=0$--5$'$, 5--10$'$, and 10--15$'$, respectively.
Each count rate is normalized by the average count rate.  From these
plots, the shape of the radial count rate profiles found to be unchanged
within 4 \% for the 1.7--7.2 keV band and 10 \% for the 7.2--12.0 keV
band.  No good correlation is seen in the 0.2--0.4 keV band and the
0.4--1.7 keV band, which is very likely due to the variation of the
non-X-ray noise and the CXB, respectively.

\section{Long-term variation of the background}

We investigated long-term variation of the background count rate.  For
the study, we used CAL CLOSED data, which is taken during the CCD is
exposed by the calibration source mounted on the filter wheel.  This
operation is performed every orbit during the satellite orbit is the
closest to the Earth and no observation is possible due to high flux of
low energy electrons.

\begin{figure}
\resizebox{\hsize}{!}{\includegraphics{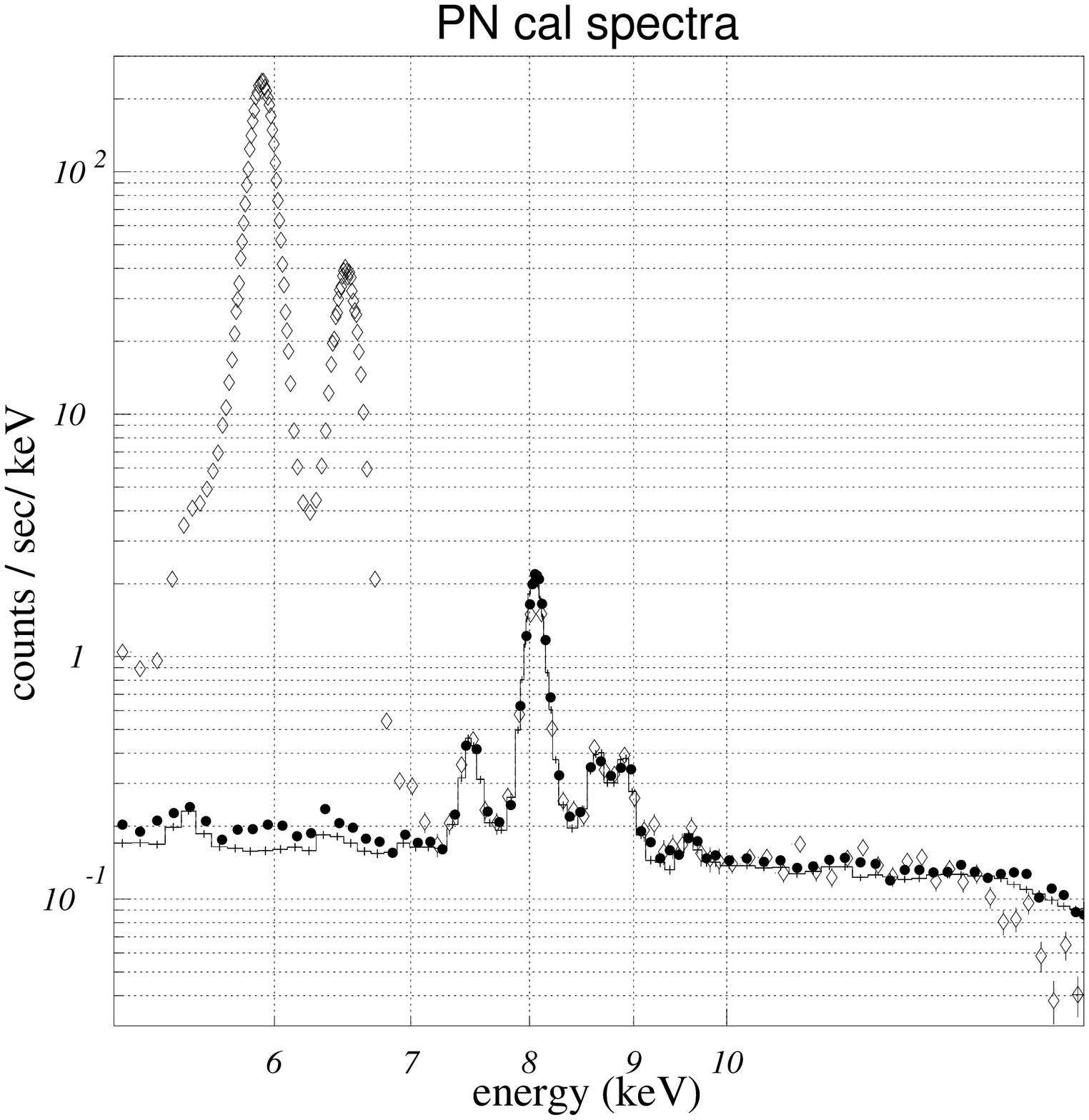}} 
\caption{Spectra of CAL CLOSED (diamond), Lockman hole (filled circle),
 and CLOSED (solid line) data.}  
\label{fig:ms_fig9}
\end{figure}

The data are free from sky X-rays and soft protons and could be used
for monitoring particle background.  Fig. \ref{fig:ms_fig9} shows the
5.0--13.0 keV spectra taken from one of CAL CLOSED data compared with the
Lockman hole data and the CLOSED data.  The contribution from the
calibration source is found only below 7.2 keV. 
Therefore, the 7.2--13.0 keV band of the CAL CLOSED data 
can be usable for monitoring the count rate of the particle origin
events.

\begin{figure}
\resizebox{\hsize}{!}{\includegraphics{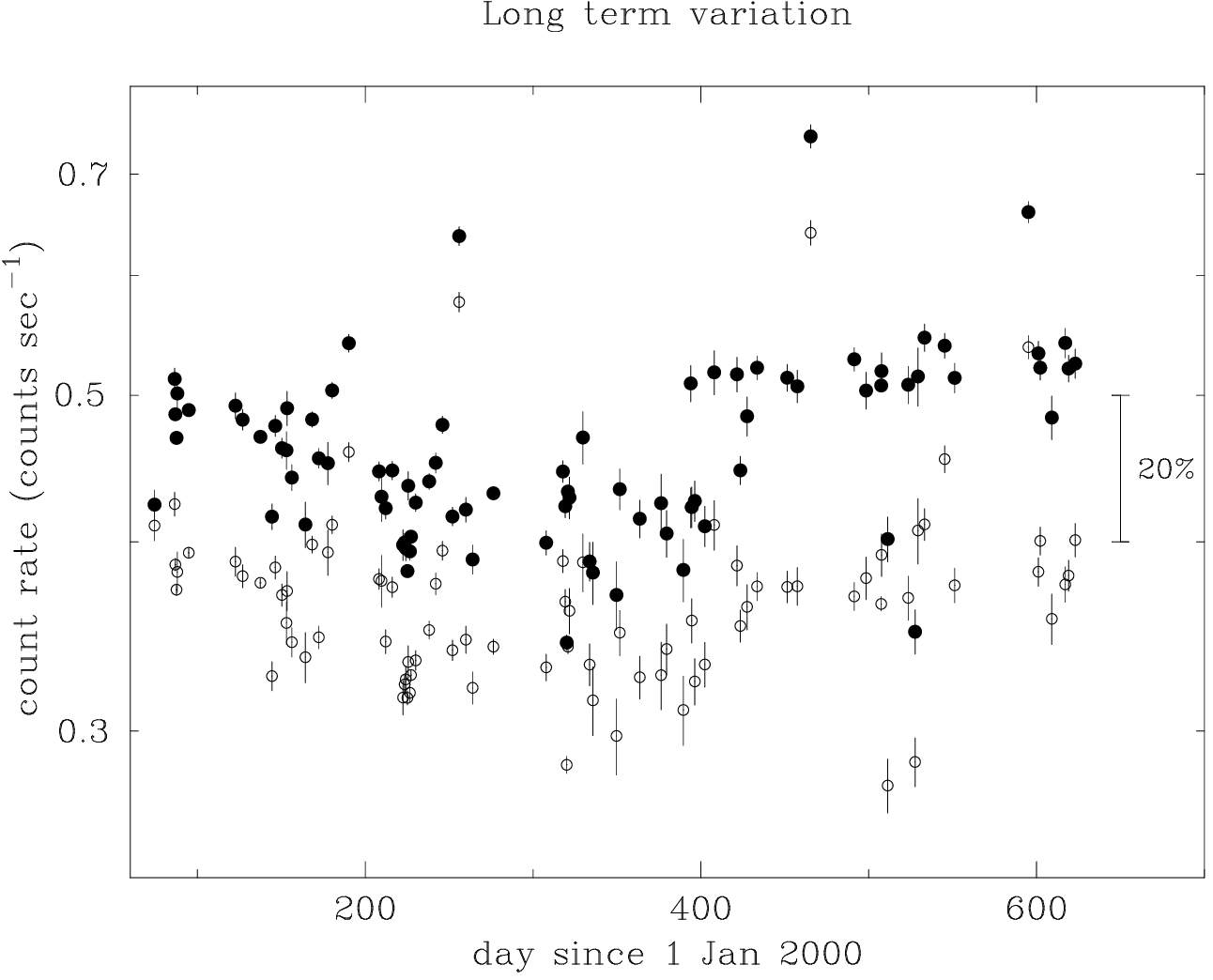}} 
\caption{ Long-term light curve in the 7.8--8.3 keV (filled circle) and
 10.0--13.0 keV (open circle) band derived from the CAL CLOSED data.} 
\label{fig:ms_fig10}
\end{figure}

Fig. \ref{fig:ms_fig10} displays the long-term light curve in the
7.8--8.3 keV and 10.0--13.0 keV band derived from the CAL CLOSED data
from March 2000 to October 2001. The average background count rate
decreased by 20 \% from March 2000 to January 2001, while, it regained
in February 2001.   Before January 2001, the same trend has been found in the Chandra ACIS
background (see Fig. 9 on Markevitch (\cite{markevitch})).
 The change at February 2001 corresponds to a change of the
operation; after February 2001,  the CAL CLOSED data are taken at positions
with high radiation, when no scientific observations can be performed.
Note that most of the data we used in Sect. 3 and 4 are taken from 103rd to 149th day,
and the long-term variations do not affect the results.

\section{Possible methods of the background modeling}

Although, the count rate of the background in quiescent period
is rather stable (within 8\% for 1 $\sigma$),
a data set may have different count rate as large as 50\% or more
from the average value as seen in the long-term light curve.
Therefore, applying the average background may not work generally.

In this section, we look for a method to predict the background count
rate for an observation data set, that should work for the data where
the source emission covers the entire FOV, like an observation of a nearby
cluster.  As a parameter for monitoring the background count rate, we
here consider the following three parameters: (1) Number of discarded
columns, (2) Count rate of 10.0--13.0 keV, and (3) Count rate of outside of
FOV.

\subsection{Number of discarded columns}

\begin{figure}
\rotatebox{270}{\resizebox{6.5cm}{!}{\includegraphics{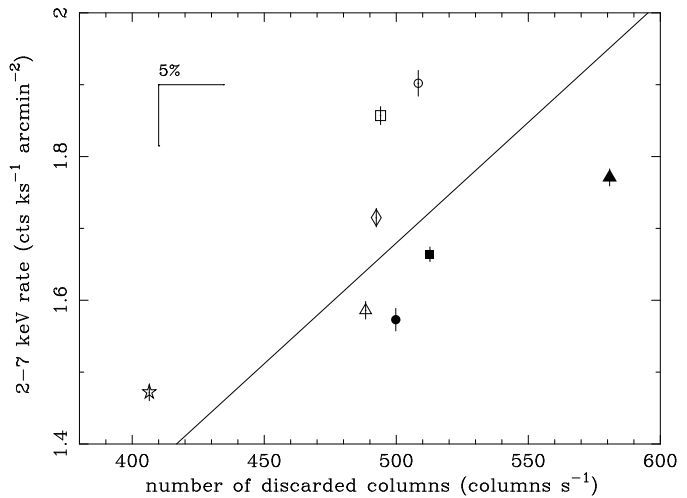}}}
\caption{Discarded column vs. 2.0--7.0 keV background count rate. The
 correlation coefficient is 0.544.
The meaning of the symbols are summarized in Table 1.
The solid line corresponds to the best fit regression line.}  
\label{fig:ms_fig11}
\end{figure}

In the data processing of EPIC-pn, the column that detects an event
above upper energy threshold and its neighbor columns are removed from
the data as a background (Appendix of Freyberg et
al. \cite{freybergb}). This is called minimum ionizing particle (MIP)
rejection. As most of these MIP events are particle events, the number
of columns that are discarded by the process is expected to be related
to the remaining background count rate. However, as shown in
Fig. \ref{fig:ms_fig11}, there is no clear correlation between the
2.0--7.0 keV count rates and the number of discarded columns. The
correlation coefficient for this data set is 0.544.

The MIP events are generated by high energy particles which pass through 
or stop in the CCD with losing its energy. 
On the other hand, the 2.0--7.0 keV background, in
particular the continuum, is attributed to a smaller number of 
particles that have a certain configuration of the energy and the
incident direction.
Therefore, the 2.0--7.0 keV background count rate would
fluctuate largely by the temporal change of the particle energy
spectrum.  



\subsection{Count rate of 10.0--13.0 keV}

For many sources, X-ray flux above 10.0 keV are negligibly small compared
to the background. Thus, the count rate above 10.0 keV could be used
for modeling the background.

\begin{figure}
\rotatebox{270}{\resizebox{6.5cm}{!}{\includegraphics{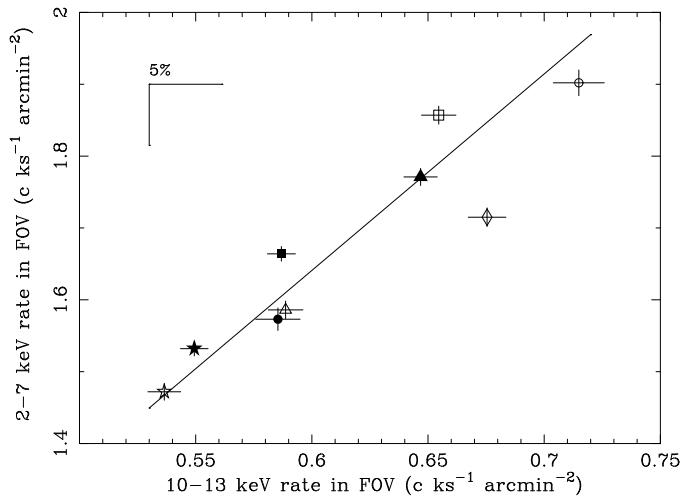}}}
\caption{Correlation between the 10.0--12.0 keV count rate and the
 2.0--7.0 keV count rate. The correlation coefficient is 0.929. The
 meaning of the symbols are summarized in Table 1.
The solid line corresponds to the best fit regression line.}  
\label{fig:ms_fig12}
\end{figure}

In Fig. \ref{fig:ms_fig12}, the background count rates in 10.0--13.0 keV
are plotted against those in 2.0--7.0 keV, showing a good correlation.
The correlation coefficient for these data is 0.929, and the scatter is
less than 8 \%.

\subsection{Count rate of outside of FOV}
 
\begin{figure}
\rotatebox{270}{\resizebox{6.5cm}{!}{\includegraphics{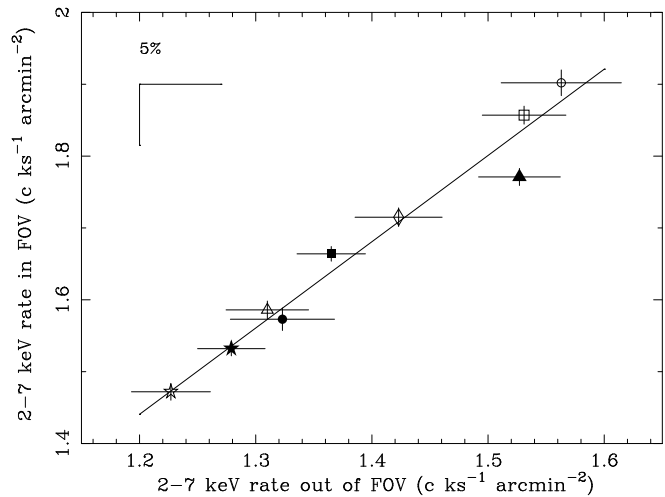}}}
\caption{Correlation between the 2.0--7.0 keV count rate of outside of FOV
 and that of in FOV. The correlation coefficient is 0.983. The meaning
 of the symbols are summarized in Table 1. 
The solid line corresponds to the best fit regression line.}  
\label{fig:ms_fig13}
\end{figure}

Events detected in the outside of FOV are mostly originated from high
energy particles and are expected to be a good monitor of the background
in the FOV.  Figure \ref{fig:ms_fig13} displays the 2.0--7.0 keV count
rate detected outside of FOV vs. that in FOV. It shows a good
correlation.  The correlation coefficient is 0.983, and the scatter is
less than 3 \%.  Therefore, the count rate of outside of FOV (hereafter
COF) shows the best correlation among the three parameters we examined
here.  Note that although the out-of-time events (Str\"{u}der et
al. \cite{struder}) that contaminates the outside of FOV have not been
subtracted, 30 \% count rate fluctuation inside FOV causes the count rate
variation in the outside FOV by at most 2 \%.

\begin{figure*}
\resizebox{\hsize}{!}{\includegraphics{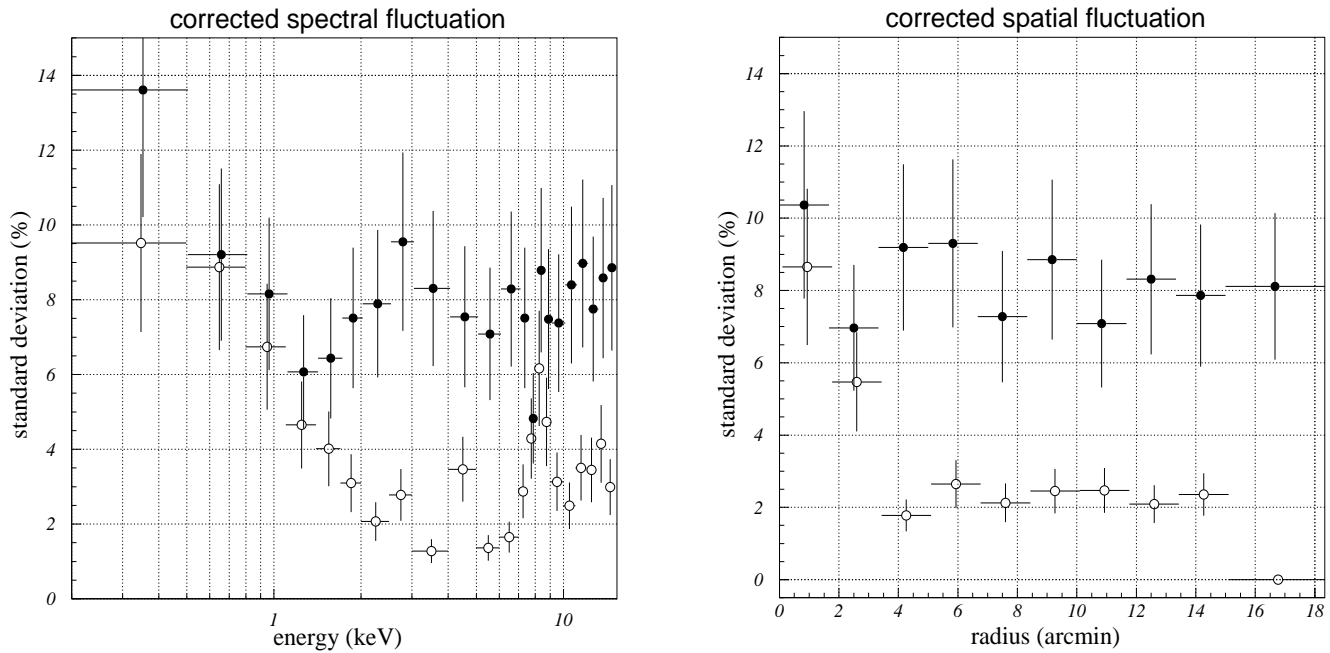}} 
\caption{Standard deviation of corrected spectra (left) and 2.0--7.0 keV
spatial distributions (right).  Filled circles and open circles
 represent data before correction and after correction, respectively. }  
\label{fig:ms_fig14}
\end{figure*}

The correlation should be used to predict background of an observation.
A predicted background of an observation can be given from the average
background scaled to have the same COF as that of the observation.  In
order to examine the efficacy, we renormalized each spectrum in
Fig. \ref{fig:ms_fig4} so that each COF in the 2.0--7.0 keV band becomes
the same as that of the average background, and calculated the
variations (standard deviations of 1 $\sigma$), which is shown in
Fig. \ref{fig:ms_fig14}.  The variation of the pn background spectra
decreases from the original scatter of 8 \% to 2 \% in the 2.0--7.0 keV,
while the soft energy band below $\sim1$ keV do not change.  This is
because the COF represents only the flux of the high energy particle
events.  The variation in the 7.0--9.0 keV band where characteristic emission
lines are detected was not improved as much as the 2.0--7.0 keV band, either.
This corresponds to the scatter seen in Fig. \ref{fig:ms_fig5} (e).  The
radial profile of the standard deviation in the 2.0--7.0 keV shows the
improvement from 8 \% to 2 \% except in the central 4 arcmin radius
region, where the contribution from the cosmic X-ray background is the
maximum.

\section{Comparison with ASCA and Chandra}

\begin{figure}
\resizebox{\hsize}{!}{\includegraphics{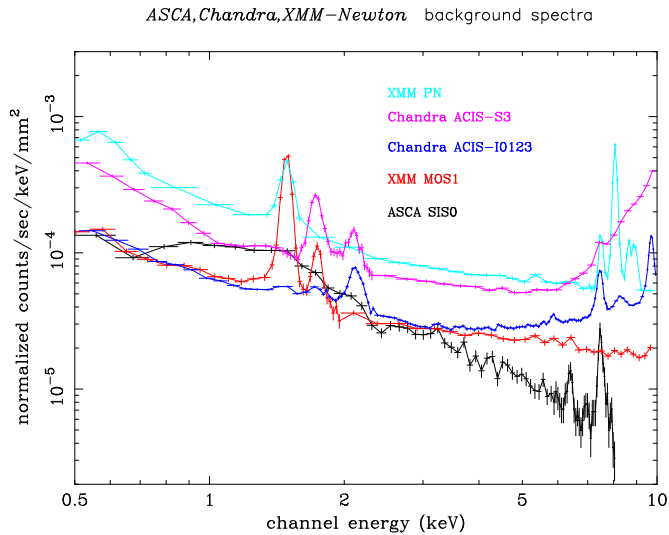}} 
\caption{XMM,Chandra, and ASCA background spectra normalized by CCD area.}
\label{fig:ms_fig15}
\end{figure}

In this section, we compare the spectra of the background of different
X-ray CCDs onboard various satellites, which include ASCA/SIS,
Chandra/ACIS-I and ACIS-S, XMM/EPIC-pn and EPIC-MOS.  For ASCA/SIS, we
used a blank-sky data base provided by NASA/GSFC, while the Chandra
ACIS-S and ACIS-I data used here are taken from blank sky observations
compiled by Markevitch (\cite{markevitch}).  Each spectrum is normalized
by the physical size of the CCD, so that they represent the background
count rate per unit area on the focal plane. The normalized spectra are
shown in Fig. \ref{fig:ms_fig15}. As these spectra include the CXB
component, a direct comparison of the particle event flux is meaningful
only above about 5keV. The difference of the background count rates
among different instruments should be explained by the
cosmic-ray-particle flux depending on the satellite orbit, and the
sensitivity of the CCDs, which is governed by the thickness of the
depletion layer and other structure (e.g. front-illuminated or
back-illuminated).

\begin{figure}
\resizebox{\hsize}{!}{\includegraphics{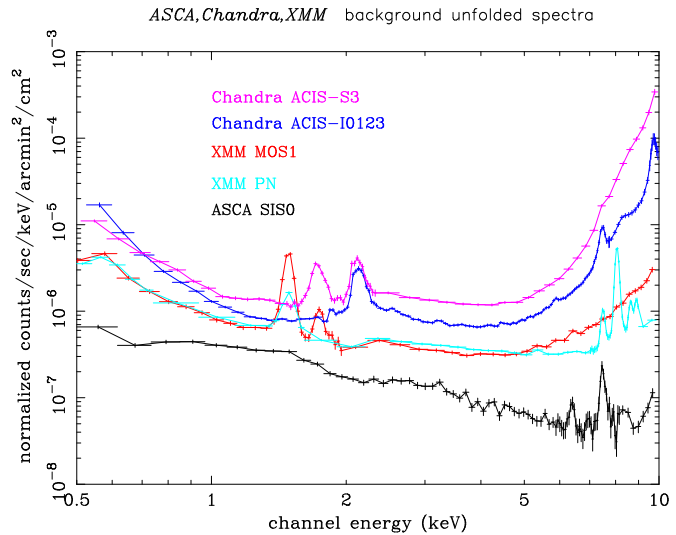}} 
\caption{XMM, Chandra, and ASCA background spectra normalized
 by effective area. }  
\label{fig:ms_fig16}
\end{figure}

In Fig. \ref{fig:ms_fig16}, each spectrum is normalized by the effective
area of the X-ray telescope plus CCD and by the solid angle of the
FOV. Thus, it represents a surface brightness of the background, and
gives a measure of sensitivity to diffuse objects.  Note that the
XMM/EPIC-pn has 10 times larger effective area but 2 times longer focal
length than the ASCA/SIS, which reduces background count rate by a factor
of ($2^2$/10) = 0.4 for the same count rates per unit area on the focal
plane.  Since the particle background count rate for  XMM is about 5
times larger than that of ASCA/SIS, the ASCA/SIS is still the most
sensitive for faint diffuse sources among those CCDs.

\section{Summary \& Discussion}

We investigated the properties of the EPIC-pn background.  A typical
light curve of an observation shows violent variabilities due to the
soft protons, while, after eliminating such high count rate periods, the
count rate is found to be rather stable.  The background energy spectrum
is consist of high energy particle events (continuum and characteristic
emission lines) dominating above 5 keV, cosmic X-ray background
dominating in the 0.4--5.0 keV band, and the non-X-ray noise contributing
only below 0.4 keV.  Background count rate from April 2000 to May 2001 is
found to be stable within 8 \% depending on energy, but it gradually
decreased by 20\% by the end of January 2001.
 In addition to this long term trend,
background count rate of an observation could fluctuate as large as 50
\% from the average.  We found that the events detected in the out of
field of view well represent the particle origin background in the field
of view as well, which can be utilized for modeling the background.  For
an actual modeling of the background, we also need take into account the
variation of the soft CXB below 2 keV from sky to sky (see e.g. Snowden
et al. \cite{snowden}).

\begin{acknowledgement}
 This work is based on observations obtained with XMM-Newton, an ESA
 science mission with instruments and contributions directly funded by
 ESA Member States and the USA (NASA). We are grateful to H. B\"ohringer
 and Y. Tanaka for useful comments. This work is supported by the
 German-Japan collaboration program funded by the Japan Society for the
 Promotion of Science and Max-Planck-Institut f\"ur extraterrestrische Physik.
\end{acknowledgement}

\clearpage

\end{document}